# Direct Imaging of Electron Orbitals with a Scanning Transmission Electron Microscope

*Ondrej Dyck[1*§], Jawaher Almutlaq[2§], David Lingerfelt[1§], Jacob L. Swett[3], Andrew R. Lupini[1], Dirk Englund[2], Stephen Jesse[1]*

[1] *Center for Nanophase Materials Sciences, Oak Ridge National Laboratory, Oak Ridge, TN, USA*

[2] *Massachusetts Institute of Technology, Cambridge, MA, USA*

[3] *Biodesign Institute, Arizona State University, Tempe, Arizona 87287*

[*]Corresponding Author e-mail: dyckoe@ornl.gov

[§]These authors contributed equally to this work

## Abstract

Recent studies of secondary electron (SE) emission in scanning transmission electron microscopes suggest that material's properties such as electrical conductivity, connectivity, and work function can be probed with atomic scale resolution using a technique known as secondary electron e-beam-induced current (SEEBIC). Here, we apply the SEEBIC imaging technique to a stacked 2D heterostructure device to reveal the spatially resolved electron orbital ionization cross sections of an encapsulated $WSe_2$ layer. We find that the double Se lattice site shows higher emission than the W site, which is at odds with first-principles modelling of ionization of an isolated $WSe_2$ cluster. These results illustrate that atomic level SEEBIC contrast within a single material is possible and that an enhanced understanding of atomic scale SE emission is required to account for the observed contrast. In turn, this suggests that subtle information about interlayer bonding and the effect on electron orbitals can be directly revealed with this technique.

Secondary electron e-beam induced current (SEEBIC) imaging using a scanning transmission electron microscope (STEM) can reveal material properties such as electrical conductivity, connectivity, and work function.[1,2] STEM-SEEBIC imaging relies on the emission of secondary electrons (SEs) from a sample induced by primary beam electrons. The accumulation of positive charge on the sample, from repeated emission of negatively charged electrons, is dissipated by grounding the sample through a transimpedance amplifier (TIA) which measures the electron current flowing into the sample. Thus, electronic properties of the sample can be explored with the resolution offered by modern STEMs. SEEBIC imaging has been used to reveal the filamentation and dielectric breakdown involved in the switching processes in valence change memory devices.[3] It has also been used to detect conductance switching in graphene nano-gaps[4] and to enable resistive contrast imaging in STEM where image intensity is directly related to sample conductivity.[1,5,6] Lattice resolution has been demonstrated in a 3D crystal to the angstrom scale[7] as well as the detection of different layer numbers of graphene.[8]

SE emission is typically described from the perspective of a macroscopic material, leveraging ensemble material properties such as dielectric function. However, these descriptions fail to describe emission (ionization) at the atomic scale due to the statistical nature of the approach. In this work, we show experimental atomically resolved SEEBIC images that exhibit contrast differences between adjacent lattice sites in the same material. To understand both the atomic resolution and variation in contrast one must abandon the macroscopic material description of SE emission, turning to first principles atomistic ionization modelling. With this framework we conclude that our image contrast is directly related to the projected electron orbital ionization cross sections. Moreover, we illustrate that a model of a simplified $WSe_2$ structure is insufficient for fully replicating the contrast observed from encapsulated $Wse_2$ leading to the conclusion that this imaging mode can be used to capture subtle changes in the electron orbital probability distributions from the effects of, in this case, interlayer interactions.

To acquire STEM-SEEBIC images an electrically conductive pathway to the device must be made and the device must reside on an electron transparent substrate (ideally suspended). To satisfy these constraints, custom devices were fabricated featuring lithographically patterned electrodes aligned with an electron transparent window after which holes were milled using a focused ion beam (FIB). A detailed report of the device fabrication process can be found in a prior publication.[9] Briefly, a 300 μm thick Si base with a 1000 nm thick thermal oxide layer and 20 nm of low-stress LPCVD $SiN_x$ formed the base substrate. Cr/Au metal electrodes were lithographically patterned and deposited on the surface to facilitate electrical connections from the STEM holder to the 2D heterostructure device. Back-side etching was performed using KOH at 80 °C to form electron transparent windows beneath the ends of the metal electrodes. Apertures were FIB milled between the electrodes to provide a region over which the 2D heterostructure would be fully suspended.

The heterostructure device was assembled following a dry transfer method. First, single-layer $WSe_2$ and single-layer graphene and few-layer hBN were mechanically exfoliated on 300 nm $SiO_2$/Si substrate with an adhesive tape. The transfer process was performed using a stacking station with a microscope and piezo controllers for sub-micron alignment. The layers were picked up using a clear polydimethylsiloxane (PDMS) stamp covered with a thin polycarbonate (PC) layer. Once all layers were on the stamp, the heterostructure was then released on the TEM chip by melting down the PC film with the proper alignment to ensure contact with the electrodes without shorting the device. The device was left overnight to avoid washing away the layers before washing off the film. The PC film was then removed in chloroform bath for 6 hours followed by a few cycles of isopropanol wash and gentle blow dry with a nitrogen gun. Cross-section and plan-view diagrams of the device structure on the $SiN_x$ window are shown in Figure 1(a) and (b) respectively. An optical image of an as-fabricated device is shown in Figure 1(c) with the exfoliated 2D flakes highlighted by the colored, dashed lines. The layer order is shown in the inset.

STEM imaging was performed using a Nion UltraSTEM 200 operated at an accelerating voltage of 100 kV and 80 kV as indicated with a nominal convergence angle of 30 mrad. Beam currents are listed with each image in the text. Electrical connections were made to the sample using a Protochips™ electrical contacting holder and the electrical leads were fed into a custom designed break-out box to facilitate making and breaking connections to the sample. A Femto DLPCA-200 transimpedance amplifier (TIA) was used for amplifying the SEEBIC signal. All SEEBIC images were acquired with a gain of $10^{11}$ V/A with a full band width of 1 kHz.

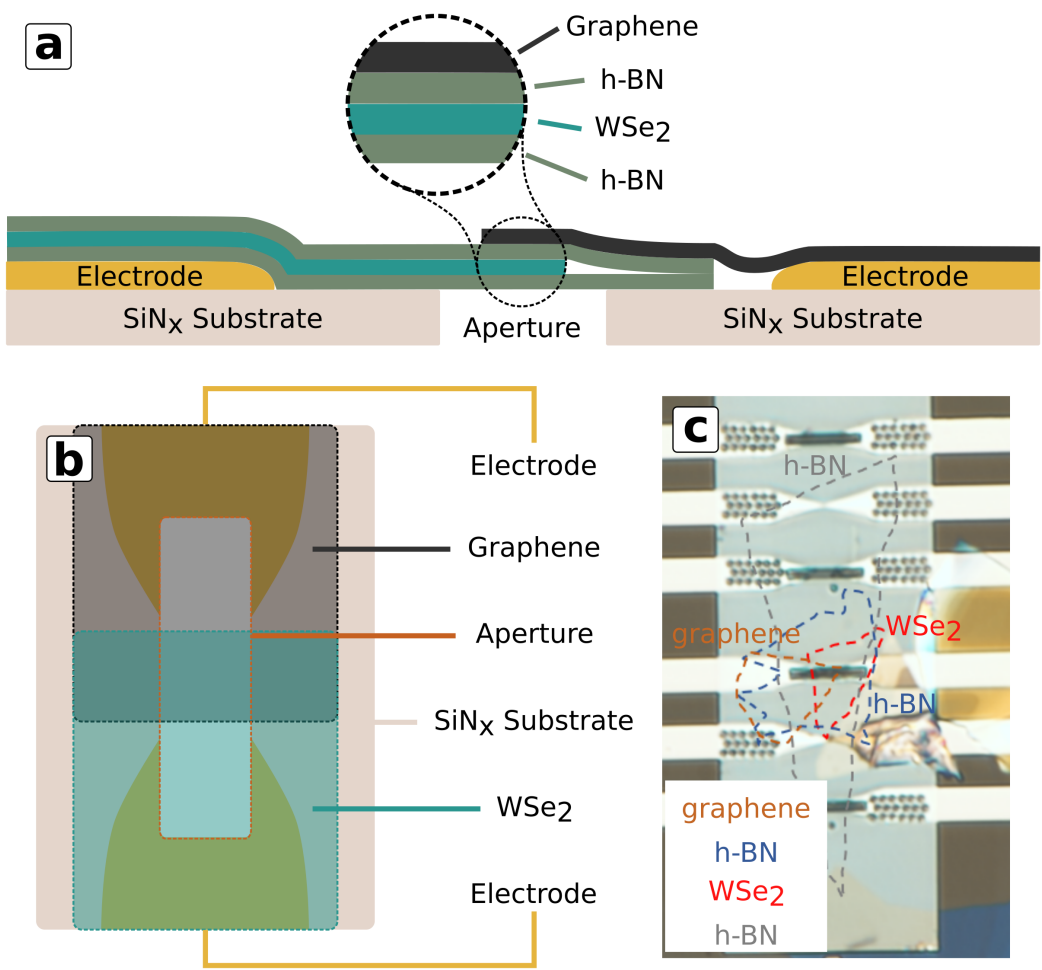

**Figure 1 Schematic and optical image of 2D heterojunction stack.** a) Cross-section schematic view of suspended device structure. b) Schematic plan view of device structure. c) Optical image of the as-fabricated device. Approximate positions of the layer locations are outlined in color coded dashed lines. Layer order is shown in the inset (substrate is on the bottom).

Figure 2 shows overview HAADF and SEEBIC images of the suspended 2D device acquired at 100 kV. In the HAADF image, Figure 2(a), we can see the end of a metal electrode and the FIB milled aperture, over which the device is suspended. Figure 2(b) shows the corresponding SEEBIC image which was acquired simultaneously with the HAADF image. Bright regions are electrically conductive and electrically connected to the TIA. This allows a clear delineation of the edge of the graphene layer (marked in the figure) which is not visible in the HAADF imaging mode. The rest of the 2D layer stack (h-BN/WSe$_2$/h-BN) appears dark, indicating it is non-conductive and/or electrically isolated from the TIA. In another region of the aperture the edge of the WSe$_2$ layer can be observed based on the Z-contrast of the HAADF signal[10], shown in Figure 2(c). In the top half of the image the WSe$_2$ layer is missing so we have two h-BN layers followed by a graphene layer on the surface. In the bottom half of the image we have our full device stack (graphene/h-BN/WSe$_2$/h-BN). The inset shows an atomically resolved image of the WSe$_2$ edge. Additionally, the h-BN can be detected using core-loss electron energy loss spectroscopic (EELS) imaging. Thus, this suite of characterization modes enables identification of each material in the stack.

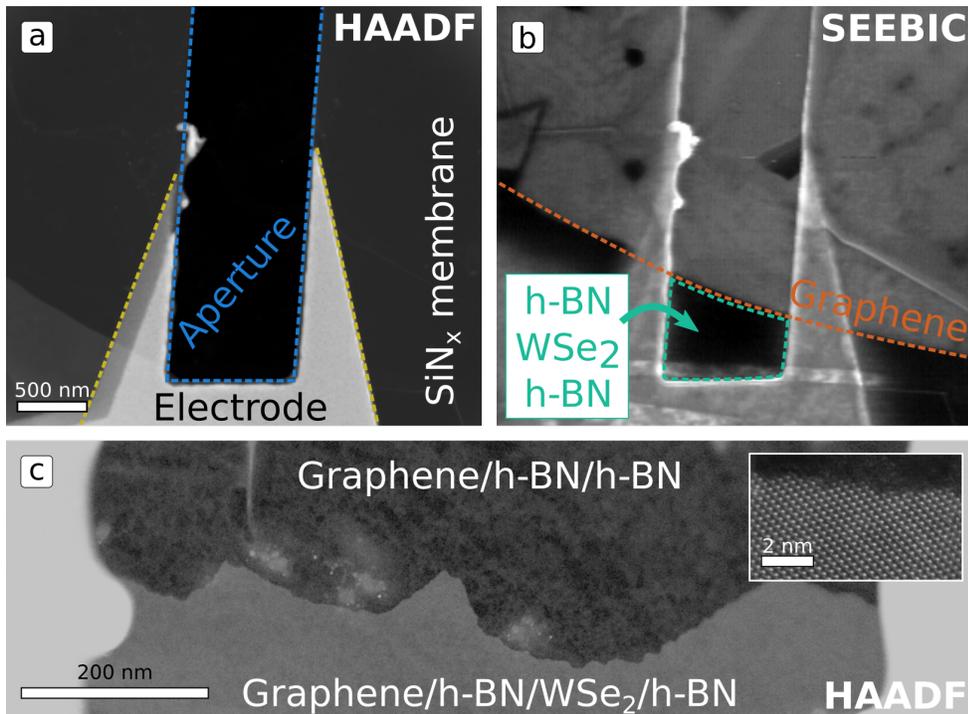

**Figure 2 Overview HAADF and SEEBIC images of the 2D heterostructure device.** a) Overview HAADF image with various major features labeled. b) Overview SEEBIC image acquired concurrently with the HAADF image shown in (a). Nominal beam current was 68 pA. The electrically conductive graphene layer can clearly be observed as brighter than the other regions. c) Overview HAADF image of the edge of the WSe$_2$ layer. The WSe$_2$ layer appears brighter due to the Z-contrast in this mode. A higher resolution image of the WSe$_2$ edge is shown inset.

In Figure 3 we examine an atomically resolved, sequentially acquired HAADF image stack with a simultaneously acquired SEEBIC image stack. Five image frames (128x128 pixels) were acquired in succession with a pixel dwell time of 8.2 ms (134 s/frame) at an accelerating voltage of 80 kV and a nominal beam current of 160 pA (1.3x10$^{11}$ electrons/frame). Single frames from the HAADF and SEEBIC channels are shown in Figure 3(a) and (b), respectively. The images were acquired on the full heterostructure stack, graphene/h-BN/WSe$_2$/h-BN, however, only the WSe$_2$ signal is evident in the HAADF signal. This is due to the much stronger electron scattering of the heavier W and Se atoms. The SEEBIC response is much more uniform and noisy, nevertheless, it appears that it also contains atomically resolved information.

There are multiple layers in the heterostructure stack so we expect that the observed response is a mixture of signals originating in each layer. These layers were not deliberately aligned with each other or with the $WSe_2$, thus creating several periodic background signals. The features contained within the SEEBIC image are therefore difficult to directly analyze. Instead we use an analysis approach that leverages the Z-contrast information contained in the HAADF image. A deep convolutional neural network (DCNN), trained to find carbon atoms in graphene[11], was used to estimate the pixel-level probability of being an atom for each HAADF image in the five frame stack. The output probability map, labeled 'DCNN' in Figure 3(c), was thresholded and the center of mass of each blob was taken as the atom locations, labeled 'Atom Finding'. Image tiles were cropped from both the HAADF and SEEBIC image stacks centered on the atom locations. A few example HAADF tiles are shown, labeled 'Tiling'. To discriminate between the W and Se-Se lattice sites, k-means clustering was performed on the HAADF image tiles using two clusters. This approach uses information contained in close proximity to the lattice site, including neighbor atom intensities and site symmetry, for a more robust decision-making feature set than that afforded by intensity alone. The k-means cluster labels for one frame are plotted as a color coded overlay on the HAADF image, labeled 'k-means Clustering'. Based on the clustering labels, histograms of individually observed atomic intensities (mean value over a 4-pixel radius from the atom position) were assembled for the HAADF stack, Figure 3(d), and SEEBIC stack, Figure 3(e). Here, we can clearly see the intensity overlap in the HAADF signal between the W and Se-Se lattice sites which are not separable using intensity alone. Finally, the mean response of W and Se-Se lattice sites for HAADF and SEEBIC is shown in Figure 3(f) and (g), respectively. These images were generated by taking the mean of all the tiles in each category. Because the tile selection has been aligned to the atomic positions of the $WSe_2$, any other features generated by the rest of the heterostructure stack (the h-BN and graphene) are smeared out into an average background. The images shown in Figure 3(g) have been rescaled using the minimum image value as a zero reference. This allows the image intensity to be interpreted directly as a current relative to the background. The colorbar lists the current values in pA. Likewise, the current values shown in Figure 3(e) have been rescaled to match. The mean values are marked with dotted vertical lines. Based on this analysis we find that the mean SEEBIC emission rate for a Se-Se lattice site was greater than for a W lattice site. A Python

Jupyter notebook that reproduces these results can be found at https://github.com/ondrejdyck/SEEBIC_electron_orbitals.

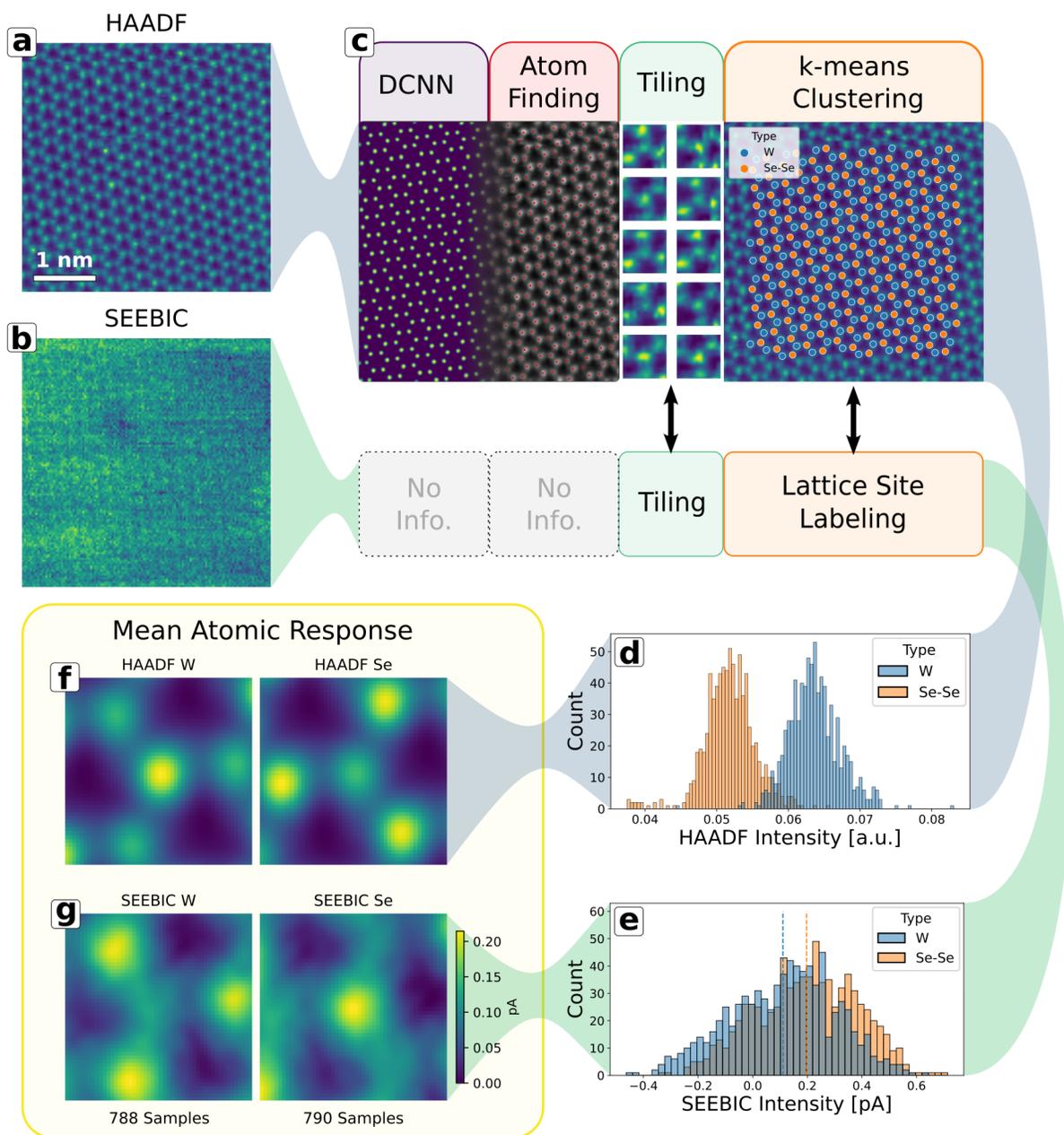

**Figure 3 Summary of image processing and intensity analysis workflow.** Single frames from a simultaneously acquired HAADF, (a), and SEEBIC, (b), image stack acquired through the full heterostructure device. (c) Depiction of image processing workflow: a DCNN was used to identify atomic positions in the HAADF image, image tiles were extracted centered on the atomic positions for both the SEEBIC and HAADF images, k-means clustering was used to identify the lattice sites using the HAADF image tiles. Histograms of the lattice site intensities for HAADF and SEEBIC are shown in (d) and (e), respectively. Histograms are color coded according to lattice site type. The mean atomic response for the HAADF and SEEBIC signals is shown in (f) and (g), respectively, separated according to lattice site type. The SEEBIC current for (g) and (e) is scaled such that the minimum response from both tiles in (g) was set to zero.

Secondary electron (SE) emission from a variety of materials and primary beam energies has been studied from a macro-scale (that is, non-atomistic) perspective.[12] These treatments leverage material properties such as work function, Fermi energy, mean free path length, etc. that consider a material as a single uniform block and were developed, generally speaking, to describe thick materials where the primary electrons (or other high energy particles) lose and transfer energy as they propagate through the material. Clearly, such descriptions begin to break down in the limit of 2D materials and with atomically resolved experiments where the emission from a single atom can be directly compared to the emission from its neighbor in the same material.

As a first approximation to the ionization rates due to irradiation at a given lattice site, one may consider total ionization cross sections calculated for the individual atoms that comprise the material. Approaches such as the binary encounter dipole approximation[13,14] have historically been used to this end, and produce atomic ionization cross sections in good agreement with those measured experimentally. While the ionization probabilities for very deep core levels in high Z elements will be essentially identical for the isolated atoms and those same atoms residing in a material, hybridization of the valence orbitals can be expected to nontrivially affect the ionization rates for any electrons in materials that are engaged in bonding. In comparison, the HAADF signal is formed mainly by beam electrons scattered to high angles by atomic nuclei and the low-loss (valence) EELS signal reports on the excitation of delocalized bosonic quasiparticles and collective electronic modes; one can expect SEEBIC signals to be uniquely capable of probing the equilibrium spatial distribution of electrons in materials. Considering only secondary electrons generated through impact scattering (and not the long-ranged dipole scattering that gives rise to excitations over longer distances) the SEEBIC intensity can be roughly understood as a measure of the local density of electronic states that is modulated by the binding energy of the electrons relative to the Fermi level and the coupling strength between the initial and final electronic states due to the electric potential associated with the incident high-energy electrons. This position-resolved electronic structure information can be viewed as a Fourier complement to the momentum-resolved electronic band structure information that is now routinely studied with angle-resolved photoelectron spectroscopies.

Since the less tightly bound electrons residing in orbitals near the Fermi energy are those most readily liberated through the electron-electron scattering, they will be the principal contributors to the secondary electron production in the limit of low beam energies. As the energy of the incident electron is increased, however, deeper valence electrons and semicore states can also be excited with reasonable probability. Indeed, the majority of the contrast observed in SEEBIC images for typical STEM beam energies (>50 keV) is concentrated in the vicinity of the nuclei.[7] As is shown in this work, however, a careful analysis allows for contrast to also be resolved in the areas between the nuclei due to the scattering from bonding electrons.

Some of the authors have previously developed computational methods for evaluating transition probabilities between the bound electronic states of materials due to the potential associated with an idealized focused electron beam within the impulse approximation.[15] Here, we build on this technique to describe the scattering-induced excitations to a continuum of unbound free electron-like states. Inspired by an approach developed for the simulation of angle-resolved photoemission spectra,[16] we have adapted these time-dependent electronic structure theory methods to allow for the escape of electrons which become unbound as a result of the beam-like perturbation by imposing absorbing boundary conditions during the electronic dynamics. In the long simulation time limit, the change in the integrated electron density resulting from the perturbation provides a measure of ionization rate associated with applying the beam-like perturbation at a particular position, which can be used to simulate the spatial variation observed in the STEM-SEEBIC intensity. Notably, screening effects are naturally incorporated in this approach at the time-dependent self consistent field level of theory. This is an important aspect of the current approach, as independent particle formalisms (which treat ionization as the removal of individual electrons from occupied orbitals) provide an unsatisfactory agreement to the observed spatial variation in secondary electron yields.[17]

Real time time-dependent density functional theory simulations were performed with absorbing boundary conditions for an isolated cluster of 2H-phase $WSe_2$ constructed from previously published crystallographic data.[18] While small clusters of these semiconducting materials will exhibit significantly blue-shifted spectral features for transitions between bound excited states relative to the bulk materials due to quantum confinement[19], these effects are not expected to drastically influence the trends we seek in the position-dependence of the probability for beam-induced ionization. All calculations were performed using a locally-modified version of the NWChem[20,21] electronic structure software, and employed the hybrid B3LYP[22–24] exchange-correlation functional and LANL2DZ[25] basis set and effective core potentials.

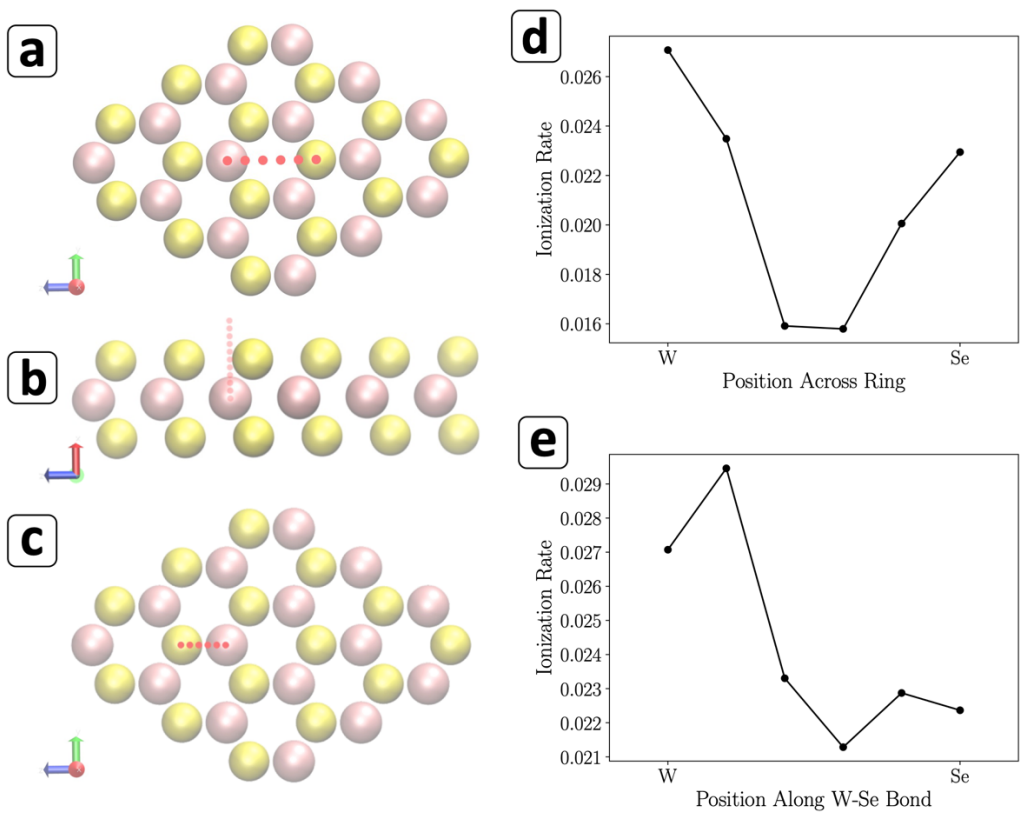

**Figure 4** Ionization rates resolved as a function of position of a beam-like perturbation from quantum electronic dynamics simulations with absorbing boundary conditions. For each of the positions indicated by red dots in the plane of the material (a) and (c), separate TD-DFT electronic dynamics simulations were carried out after applying an electric potential impulse associated with the presence of an electron at each of the 10 points normal to the plane of the material as shown for one of these in-plane positions in (b). Points in the out-of-plane direction (b) were chosen to be equally spaced relative to both W and Se atoms' out-of-plane positions. The sum of the total charge lost to the absorbing boundary conditions for each of the out-of-plane positions is reported as the 'ionization rate' for each of the in-plane positions along the W-Se bonding direction (e) and across one of the "rings" (d).

Due to contributions to the overall ionization rate from dipole-allowed transitions to unbound and metastable, auto-ionizing states, the SEEBIC signal can be expected to show a significant delocalized component. For example, even for perturbations applied near the center of the "ring" (i.e. where very little electron density resides) the calculated rate of ionization is still over half that associated with the highest rate observed along the W-Se bonding direction. While this delocalized response is responsible for a relatively position-independent background contribution to the SEEBIC signal, sharply position-dependent features will be contributed by electronic transitions that are promoted through quadrupole and higher-order terms in the multipole expansion of the external Coulombic potential that are only active over short distances (i.e. through "impact scattering"). The ionization rate will also carry some position dependence for the dipole-allowed transition rates through modulations of the intensity of the electric field emanated by the external electron over the volume of the transition density, as even the electric dipole contribution to the matter-field interactions decays quadratically with distance from the field source. While the simulated valence ionization rates do exhibit position-dependent variations, there are notable discrepancies in the predicted rates and those observed in the SEEBIC images. This result suggests that to properly account for the observed intensity the first-order approximation of an isolated cluster of the $WSe_2$ is insufficient. This is understandable, as the SEEBIC images were collected for $WSe_2$ laminated in multiple layers of materials with different dielectric properties, where the atomistic details of the interfaces nontrivially alter the SEEBIC intensity relative to the ionization rates of the isolated $WSe_2$ monolayer.

Together, these results suggest that direct imaging of the spatial distribution of electron orbital cross sections via SEEBIC has the potential to reveal nuanced information about the local electron distribution and bonding. Coupled with techniques such as 4D STEM and EELS, which can both be acquired in parallel with SEEBIC, this technique promises to provide powerful insight into materials chemistry. The TD-DFT modelling approach presented here indicates the direction future investigations must take to fully account for SEEBIC contrast on the atomic scale. That the simplified model only partially accounts for the observed contrast in our experiments suggests that this endeavor will be rich with new information.

**Acknowledgement**

This work was supported by the U.S. Department of Energy, Office of Science, Basic Energy Sciences, Materials Sciences and Engineering Division (O.D. A.R.L., S.J.), and was performed at the Center for Nanophase Materials Sciences (CNMS), a U.S. Department of Energy, Office of Science User Facility. J.A. acknowledges the fund from KACST-MIT Ibn Khaldun Fellowship for Saudi Arabian Women.